**Health Advertising on Facebook: Privacy & Policy Considerations**
Authors: Andrea Downing, Eric Perakslis

**Abstract**

Digital medicine technologies offer convenience to consumers and can improve outcomes for chronically ill patients navigating their health journeys. An essential element for patients to adopt new health technologies is trust. By providing digital tools designed to improve health, sometimes the most sensitive personal and medical information is shared. Digital medicine apps and services must maintain that trust by ensuring that all uses, and privacy of personal health data are protected across the complex technology ecosystem that these companies utilize.

In this study, we analyzed content and marketing tactics of digital medicine companies (n=5) to evaluate varying types of cross-site tracking middleware (n=27) used to extract health information from users. More specifically we examine how browsing data can be exchanged between digital medicine companies and Facebook for advertising and lead generation purposes. The examples we analyzed, focused on a small ecosystem of companies offering services to patients in the cancer community who frequently engage on social media. We co-produced this study with public cancer advocates leading or participating in breast cancer groups on Facebook.

Some companies in our content analysis may fit the legal definition of a Personal Health Record Vendor covered by the Federal Trade Commission, others are HIPAA-Covered entities. Following our analysis, we raise policy questions about what constitutes a data breach under the Federal Trade Commission's Health Breach Notification Rule. A few examples show serious problems with inconsistent privacy practices, and reveal how digital medicine dark patterns may elicit lead data from both patients and companies serving ads. Further we discuss how these common marketing practices enable surveillance and targeting of medical ads to vulnerable patient populations, which may not be apparent to companies targeting ads.

**Introduction**

A "Dark pattern" is a user interface design which benefits an online service by nudging, coercing, or deceiving users into making unintended and potentially harmful decisions (1). "Privacy Zuckering" is a known type of dark pattern termed originally by Tim Jones at Electronic Frontier Foundation in 2010, and happens when a user is tricked into publicly sharing more information than a user really intended to share (2). When this specific type of dark pattern is employed to elicit public data from patient populations online, one might consider the sensitivity of health data involved. In the field of health privacy and cybersecurity, when protected health information (PHI) is leaked or stolen, the potential harms include physical harm, economic harm, psychological harm, reputational harm and societal harm (3,4).

With the explosion of social media over the past decade, online communities of patients exist on social media platforms. Health and pharmaceutical companies spent almost $1 billion on just Facebook mobile ads in 2019 (5).  Patient communities on social media who have shared knowledge about their health have generated increasingly large marketing channels for digital medicine and pharma companies to target ads to patient populations (6).

Our study focused on a small ecosystem of cancer patients who are regularly navigating between Facebook and digital medicine, personal health records, and diagnostic testing services.  Social media platforms like Facebook have become common places for patients to seek support from their peers online, while social media is filled with ads relating to health conditions (7). We focus solely on Facebook's ad model in this analysis, and may broaden to other social media platforms in future studies.  While this analysis does not tie specific harms to dark patterns utilized, known examples of harm when publicly exposing health information can include risk of discrimination, psychological harm, and exposure to fraud or scams based on the health information shared.

We chose to focus on cross-site tracking middleware used by digital medicine companies because these tools may make patient populations vulnerable to online scams, medical misinformation, and privacy breaches (3).  Patient communities' digital footprint expanded exponentially when patients turned to social networks such as Facebook seeking support, knowledge and advice during a health diagnosis (6).  Genetic testing companies, health services, and patient communities can become unwitting participants in digital dark patterns when posting or engaging with ads on Facebook.

We analyzed cancer related health companies (n=5) using third party cross site tracking tools to patients' behavior between their own websites and Facebook.  This process identified how companies are able to leverage Facebook's health-related ad targeting tools to generate data about cancer patient advocates on social media as marketing leads. Our analysis showed 3 of 5 companies where neither the vendors nor Facebook were compliant with their own policies or claims about privacy.  2 of the 5 companies, Ciitizen and Invitae, targeted ads that were consistent with their privacy policies.  Yet, all companies in our analysis created digital footprints to enable ongoing tracking and surveillance of patient populations on Facebook. Findings of non-compliance were similar in a recent cross-sectional study demonstrating that in an ecosystem of medical and digital health or digital medicine apps available on Google Play, only 47% user data transmissions complied with each company's own privacy policies (7).

The scope of this study focused on examples which may fit the definition of Federal Trade Commission's PHR Vendor (8), and also examples of CLIA certified diagnostic testing laboratories that are likely HIPAA-covered entities (9).  Some examples deal with protected health information (PHI) if they are HIPAA-covered entities (10.)  Other examples may qualify PHR identifiable health information, if covered by the Health Breach Notification Rule.  By following the reproduction steps, the common theme in examples we identify shows how each company tracks patients off-Facebook while targeting ads to reach patients on Facebook.  3 of the 5 companies went further to re-identify patients as leads using common advertising tools.

Companies involved may also be unwitting participants exposing more about the patient populations they serve through Facebook by creating rich digital footprints of patient populations who interact with their ads and services.

Given the known limitations and potential workarounds for the HIPAA Privacy and Security Rules, patients are mostly on their own with respect to understanding how companies utilize their personal and health data, especially when asking questions about their health conditions on social media.  Despite the known gaps in regulation, end user license agreements remain the standard for eliciting consent for data use with most digital toolsets and platforms and patients are typically on their own in assessing the risks of using digital toolsets (11). While utilizing similar marketing practices, the line between "market" research and ethical human subjects research remains unclear when patient populations lack comprehensive privacy regulation in the United States .

**Methods**

Public patient advocates in the hereditary cancer community (N=20) were invited to participate in co-production of this research at a response rate of 50% (N=10).  These patient advocates include a small sampling of the broader population.  Specifically, public metadata for hereditary cancer communities on Facebook consists of about 73 Groups ranging in size from 36 to 13,000 people (12).  Out of this population 3 of the 10 participants were active admins of at least one Facebook support group for breast cancer.  All participants were a member of at least one breast cancer support group over a time period between 2008-2021.

Facebook has a tool in user settings that allows users to see companies tracking browsing data in "Off Facebook Activity," which can also be downloaded into an archive of JSON files (13).  The patient advocates who co-produced our data (N=10) were asked to download their full Facebook archives as JSON files. Participants could also look at the data via Facebook's user interface using "Off Facebook Activity" in their user settings and either provide screenshots. Each participant checked if they found digital medicine apps in their "Off Facebook Activity" JSON files and then verified whether these users of PHR Vendors or HIPAA-Covered Entities had authorized access to their data.  As a next step we analyzed each digital medicine company's cross-site tracking tools and compared tools they used with their Privacy Policies.  As a final step, we checked Facebook's Ad Library to identify types of ads being run by each company. We also examined how each ad's URL exposed information passed data from Facebook to third parties.  From the 5 companies we identified in JSON files,  we identified 27 third party CDN or cross-site tracking tools (Figure 1).

Our method was to take the following reproduction steps to identify digital medicine companies that may be tracking users across websites.  These participants took the following steps:

1. Download the full archive history of "My Facebook Information."

2. For each of their archives, we asked participants to specifically review the audit log of your_off-facebook_activity.json
3. Within this JSON file, we asked participants to provide the list of health apps that appeared in their history
4. We then took this list to check each Vendor's website for 3rd Party Ad Trackers. This can be done with any basic CDN Tracker such as EFF's Privacy Badger ([https://privacybadger.org](https://privacybadger.org)), or The Markup's Black Light Tool ([https://themarkup.org/blacklight](https://themarkup.org/blacklight)) (14,15)
5. If 3rd Party ad trackers were found, we checked the Privacy policy to see what was disclosed to users, and whether that matched what we found.
6. We checked Facebook's disclosures to users about how PHI is shared, and how it is used.
7. As a final step, we checked Facebook's Ad Library to check each PHR History and Types of Ads posted. We examined each ad's URL to identify FBCLID and use of Crowdtangle.

**Coordinated Disclosure & Timeline**

While the CDN and cross-site tracking tools utilized in our study are commonly used in digital advertising, 3 of the 5 digital medicine companies in our analysis used 3rd party tools to re-identify or re-target users as marketing leads without clear language in their privacy policies or authorization from users. Given our analysis uncovered examples that may qualify as a breach of personally identifiable information either under HIPAA or the Health Breach Notification Rule, a crucial step to our research was a coordinated disclosure with companies involved.

According to CERT.org, coordinated disclosure of a vulnerability requires multiple stakeholders to analyze a vulnerability to be able to disclose it to the public and provide guidance on how to mitigate or fix it (16). Vulnerabilities and disclosures to impacted parties often follow unique paths where no two disclosures are alike. Through the process of coordinated disclosure, our goal has been to work in good faith with various stakeholders and make sure the vulnerability is addressed accordingly and that the correct information reaches the public.

When attempting to locate coordinated disclosure policies for some of the impacted companies, we were unable to find points of contact at 2 of the 5 companies to coordinate our findings. The companies we analyzed only represented a small sampling of PHR Vendors or genetic testing companies, and it became apparent we would not be feasible to reach out to thousands of other potentially impacted companies who target ads. Therefore, we reached out to Cert.org to assist with multi-party disclosure to impacted vendors. Cert.org provided guidance and to navigate coordinated disclosure to impacted companies.

Cert.org provided assistance to ensure we were following proper guidance for coordinated disclosure in 2021.

- November 15, 2021:  Attempted to locate points of contact for disclosure at each company, but realized direct coordination would not be possible.[1]
- November 24, 2021:  Disclosure to BioISAC
- December 1, 2021:  Disclosure to Cert.org
- December 10, 2021:  Disclosure to Ciitizen & Invitae
- December 13, 2021:  Disclosure to Color Genomics (no response)
- December 16, 2021:  Submitted report to FTC
- December 17th:  Disclosure to Cert.org to request help with multi-party disclosure.
- December 31st:  1 of 5 vendors (Health Union) updated their privacy policies.
- January 2022:  Invitae & Ciitizen response to disclosure.
- January 2022:  Facebook removes all sensitive health ad targeting endpoints (19)

Cert.org solicits and posts authenticated vendor statements and references relevant vendor information in vulnerability notes.  While one company requested them, we did not provide JSON files of participants to the companies involved, in order to protect the privacy of participants.

During our coordinated disclosure, it is notable that Ciitizen and Invitae were the only examples that responded to our report.  While both examples showed content views and page views in participants' JSON Files, Ciitizen and Invitae responded by removing all social media ad targeting tools to further assess the impact of these tools on their users.

**Results**

Through this study we were able to demonstrate and reproduce how sensitive data flows from our examples of digital medicine vendors to Facebook through cross-site trackers or content delivery networks (CDN's).  Some of those same vendors (2 out of 5) target ads on Facebook to share data back with digital medicine vendors and elicit more data from users as leads.  Trackers and types of CDN's passing data between Facebook and digital medicine Vendors varied widely.   The common user experience may be described as follows, though steps may vary by vendor and user (Figure 2):

> Step 1: User signs up for digital medicine app or genetic testing and agrees to the Companies Terms of Service.

---

[1] Disclosure was not possible for some parties if their websites did not provide a coordinated disclosure policy or a security point of contact. Further, this study only analyzed a small sampling of 5 companies, where impacted health vendors using these practices number in the thousands.

Step 1a:  Separately the user creates an account on Facebook, or has an established account.

Step 2: Vendors Embed 3rd Party Tracker in PHR Website

Step 3:  Multiple Trackers Share "Off Facebook Activity"

Step 4:  Off Facebook Activity from the Vendor updates User "Ad Interests" algorithms on Facebook

Step 5:  Facebook's predictive algorithms begin to promote health-related ads to the user based on health interests.

Step 6:  The Vendor targets ads to users with specific health interests, and in some cases use quizzes or sign-up forms to enrich their lead data.  Lead data is passed from Facebook to the Vendor's CRM system.

While these steps vary in detail between participants who shared their "Off Facebook Tracking", the remainder of this report will detail the examples of vendors where we have examples of Off Facebook Activity, and ad targeting that did not match privacy policies and/or user settings.  While Facebook does disclose how their proprietary algorithms predict health interests about a user, we do know that ad targeting for certain diseases such as cancer are available based on data collected from Facebook about users' web browsing behavior.

We outline specific examples where "Off Facebook Tracking" showed examples of digital medicine companies in JSON archives of participants.  While this is a small sampling of a larger population and digital ecosystem, these examples highlight how cross-site tracking works when users navigate between PHR's and social media, while being tracked through Content Delivery Networks.

Example #1:  Color Genomics

One participant identified Color Genomics in their "Off Facebook Activity" JSON files.  Color Genomics provides a DNA health report that analyzes up to 74 genes that fall into 3 categories: 30 genes that impact risk for breast (including the breast cancer genes BRCA1 and BRCA2), ovarian, uterine, colon, melanoma, pancreatic, stomach, and prostate cancers.  Color Genomics is a CLIA certified lab and a HIPAA-Covered entity.

With respect to privacy practices, Color states that they require a user's authorization before disclosing PHI for marketing purposes in a notice dated May 25, 2018 at the time of our study.  They represent to users that they do not rent, sell, or otherwise use patient data (Figure 8).  One participant noted their cookies on Color's website are turned Off, which indicates that users did not authorize sharing of information for advertising purposes.  Using the reproduction steps outlined in our methods section, we identified the 3 cross-site trackers (Figure 1).

Notably, Color used Leadfeeder is a marketing solution that enables companies to re-identify leads based on their visits to a website, and enrich the data without explicit consent from users (Figure 3) (17). From the Facebook Ad side, we also looked at trackers being used by Color when users click on their ads.  In one ad, the "Shop Now" Button goes to the following URL providing data to a social advertising service called Nanigans.  (Figure 4) (18).  Here is one example of an Ad link passing from Facebook to Nanigans, where we have redacted ID's exposed in URL's:

> http://api.nanigans.com/target.php?app_id=[redacted]&nan_pid=[redacted]&target=https%3A%2F%2Fhome.color.com%2Ft%2Fstart%3Futm_source%3DFacebook%26utm_campaign%3DProspecting_OC%26utm_medium%3DROF%26utm_content%3Dvideo%26code%3DVAF6OM0JMH%26%26nan_pid%[redacted]%26ad_id%[redacted]

As a third-party marketing service, Nanigans was acquired by a data analytics company called Sprinklr in 2019 (19).  Sprinkler provides a service to companies called Unified-CXM which brings together user interests across platforms. For example, Unified CXM works across platforms to re-identify and join data about users across social media platforms (Figure 5) (20). We did not have visibility into Color's specific use of Nanigans, nor did we receive a response during our disclosure process.  The history of ads that we originally analyzed were removed from Facebook's Ad Library during the disclosure process, sometime before December 2021.

Example #2: MySupport360 & HereditaryCancerQuiz.com

One participant identified Mysupport360.com and HereditaryCancerQuiz.com in their JSON files.  These services are patient-facing hereditary cancer awareness campaigns created by Myriad Genetics website that focuses on three of the company's genetic tests.  Myriad Genetics is a CLIA certified lab that provides diagnostic testing for cancer genetics.  Myriad's tests provide patients and healthcare providers with insights into inherited cancer risk and molecular features of cancer tissues, and also has genetic tests related to psychiatric drugs and other conditions. As a covered entity under HIPAA, Myriad's lab, MySupport360, and HereditaryCancerQuiz.com may or may not be covered by the FTC's breach notification rules. The MySupport360 site specifically provides consumers with tools to find genetic tests for BRCA and other genes, without engaging directly with a physician, yet they are a diagnostic lab covered by HIPAA.

Through our analysis from participants we identified 2 instances of Off-Facebook Activity as late as June 24, 2021 in participant JSON files.  The user did not provide written authorization to MySupport360, HereditaryCancerQuiz.com, or Myriad Genetics to disclose any information to Facebook for marketing, ad targeting, or any other purpose.  Myriad, the parent company of MySupport360, showed targeted ads on Facebook in the form of a 'Hereditary Cancer Quiz' to gather personal details about a person's health and family history.  It is not disclosed to users that PHI for hereditary cancer entered in the 'quiz' or input form will be used as lead information for Myriad.

The ads we originally analyzed from Facebook's Ad Library have since been removed without explanation. These ads passed a link referral through a service called Crowdtangle (see Crowdtangle section for more information). This example came from a specific user clicking on the hereditary cancer quiz link in the following format:

https://www.hereditarycancerquiz.com/?fbclid=[redacted]_0AJ_[redacted].

Before ads were removed, when a user clicked on the ad from Facebook to Myriad's quiz, a parameter called "FBCLID" was used as shown in the example above. FBCLID stands for Facebook Click Identifier (21). Since Mid-October 2018, FBCLID has been appended to all outgoing links in Facebook.

The SDK documentation from Facebook on FBCLID indicates that data is gathered about the user when clicking on an ad. For example, once the FBCLID ID is created, passed to Crowdtangle, the landing page quiz appears as if it's a public health service, this 'health quiz' passes personal health information to both Facebook and Myriad Genetics as lead information.

Information Collected in the quiz included:

- Date of Birth
- Sex
- Are you of Ashkenazi Jewish Ancestry?

HereditaryCancerQuiz.com also included some of the more invasive trackers out of the examples we identified, and posted similar ads by Myriad Genetics. For example, HereditaryCancerQuiz.com was the only example we identified with Facebook Pixel installed directly on their website (Figure 1). Further, we noted one custom field being shared between Myriad and Facebook. In our Policy Questions section, we draw comparison to the way that Flo Fertility had shared custom fields with Facebook. Given custom fields had been created for users, it would be helpful for Myriad to disclose what type of data had been shared.

Example #3: Invitae

Invitae is a CLIA Certified diagnostic testing lab that offers clinical genetic tests. The company states: *Invitae's mission is to bring comprehensive genetic information into mainstream medical practice to improve the quality of healthcare for billions of people. From day one, patients owning and controlling their genetic data has been one of our core principles.*

Two participants showed Invitae in their JSON files in the form of content views and page views on their website (Figure 9). However Invitae is perhaps the most benign example in this report in contrast to the others. Invitae transparently discloses that they use cookies, and clearly states that they share information with advertisers. Invitae is a CLIA Certified diagnostic testing company. Invitae's Privacy Practices clearly outline how cookies are used. Their cookie policy states:

> We use other tracking technologies similar to cookies, such as flash cookies, web beacons, or pixels. These technologies also help us understand how you use our Services in the following ways:
> - "Flash Cookies" (also known as "Local Shared Objects" or "LSOs") to collect information about your use of our Services. Flash cookies are commonly used for advertisements and videos.
> - "Web Beacons" (also known as "clear gifs") are tiny graphics with a unique identifier, similar in function to cookies. Web beacons are embedded invisibly on web pages and do not store information on your device like cookies. We use web beacons to help us better manage content on our Services and other similar reasons to cookies.
> - "Pixels" track your interactions with our Services. We often use pixels in combination with cookies.
>
> We generally refer to cookies, web beacons, flash cookies, and pixels as "cookies" in this Policy.

At the time of this study, we identified 5 cross-site trackers on Invitae's main website (Figure 1). The majority of these middleware tools used cookies to improve operation of Invitae's site, not to specifically gather leads or re-identify website visitors as leads, as shown with the example from Color Genomics. Invitae does not show ads targeted to users passing through Crowdtangle. While page views passed to Facebook do not re-identify patients as leads, or pass custom fields to Facebook, any users clicking on Facebook ads are then sharing information with Facebook about their health interests. These "ad interests" are then available for other companies to retarget the patient on Facebook.

Example #4: Health Union

Participants identified ads for AdvanceBreastCancer.net in their social media feeds, but did not find cross-site activity in their JSON files. These ads were run by a digital medicine service called Health Union. Health Union states their online health communities provide support, information and a sense of connection across a variety of chronic health conditions in oncology, immunology, neurodegenerative, genetic, and general medicine.

Health Union's business solutions are targeted to pharma, marketing research, and clinical trial services as their primary customer to "enable companies to connect and engage in transparent ways with highly qualified people who interact with our communities." AdvancedBreastCancer.net is one of several online communities run by Health Union. We identified 5 cross-site trackers or CDN's in our analysis of Advanced (Figure 1).

Notably, we saw a disconnect between Health Union's claims about privacy and their activities. The company claimed on their main page that they never sell health information without "specific permission to do so." (Figure 6). However, Health Union's privacy policy stated that users must submit a request to opt out the sale of information to privacy@health-union.com or visit the Walt Disney Company's privacy rights page (Figure 7). While Health Union did not respond to us directly during the coordinated disclosure process with Cert.org, the company updated their privacy policy on December 28, 2021.

Example #5: Ciitizen

Ciitizen is a service that enables patients to organize health records from multiple sources, such as different EHR Systems (22). Ciitizen was acquired by Invitae, example #3, in 2021 (23). This is one of the two more benign examples we analyze in our report. Ciitizen disclosed in their Privacy Policy how they use cookies, and provided statements about how they share information with advertisers. No custom fields were identified in the JSON files of the participants in our study, only views of web pages and content on Ciitizen's website. The data passed back to Facebook is shown only as content views on Ciitizen's service which include the dates that a patient viewed content on Ciitizen's site. The ads we reviewed for Ciitizen in Facebook's Ad Library have since been removed from Facebook's Ad Library and from JSON files. Yet, any ads created by Ciitizen and other examples in this study feed predictive algorithms to Facebook through cross-site tracking, which enables retargeting of patients based on their health "interests." Notably, Ciitizen provided the most comprehensive response to our initial report during coordinated disclosure. In direct response to revelations in our report, both Ciitizen and Invitae reported back that they took down Facebook's ad tools. They are further assessing the impact of ad targeting tools they are using on other social media sites.

Policy Questions

At the time of this study, the FTC has not once enforced the Health Breach Notification Rule since its creation in 2009 (8). It would help for policy makers and digital medicine companies alike to evaluate these real-world examples that may apply to enforcement of the FTC's Health Breach Notification Rule. There are a range of questions to consider. Do some of the examples in this study fit the definitions in the Health Breach Notification Rule? To what extent information "managed, shared, and controlled by or primarily for the individual '' fit legal definitions in this rule for different types of companies outlined. Are general statements about marketing practices in privacy policies sufficient "disclosure" for purposes of the health breach notification rule? Given the sensitive health information exchanged between

Facebook and HIPAA-Covered entities, does Facebook itself fit the legal definition of a Vendor of Personal Health Records?  Where do we draw ethical lines between market research and medical / human subjects research in this exploding digital medicine ecosystem?  What constitutes a"breach" if patients data are combined with existing information on other platforms or if patients are tracked across platforms?

We hope our analysis, which was co-produced by advocates affected by these questions, provides helpful insight on real world data from the perspective of these patient populations.  Notably, Facebook announced in November 2021 that they would be removing all detailed ad targeting endpoints for sensitive health information (24).

A recent FTC Settlement with a digital medicine app called Flo may serve as a legal precedent for the comparisons in our study (25). In this FTC Settlement, Flo handed users' health information out to numerous third parties, including Google, Facebook, marketing firm AppsFlyer, and analytics firm Flurry, Inc (25). Notably, the Flo Settlement did not include a violation of the Health Breach Notification rule, but rather was focused on deceptive statements to users, per Section V of the FTC Act. One of the key activities from Flo's case was using CDN trackers to pass custom fields about users menstrual cycles directly to Facebook using custom fields.  Based on our understanding of the Health Breach Notification Rule, examples we share in this report must meet the following criteria to qualify as a "breach"(8):

1. The business or service must "offer or maintain a personal health record.
2. The PHR Vendor's Terms of Service and Privacy Disclosures must fail to disclose sharing of personally identifiable information with 3rd parties.
3. If unauthorized access to PHR identifiable health information occurs, the PHR Vendor must notify users of a breach.

Further research is necessary to enumerate ways that populations on the platform may have been targeted by malicious actors, scams, and medical misinformation through the same tools and methods we've identified.  It would be beneficial to investigate how these middleware options can be utilized to discriminate against patients, and potentially target large scale populations with medical misinformation.  Upon discovering health apps using cross-site tracking, some of the advocates who reported their findings that they felt "duped" "exploited" and "violated" after seeing ways that digital medicine services and genetic testing companies were tracking the cancer community's Off Facebook Activity.

## Conclusion

Health privacy is a basic requirement in digital medicine for reducing the abuse of power and supporting patient autonomy (4).  We demonstrated that personal data and personal health data can be easily obtained without the aid of highly sophisticated cyberattack techniques, but rather commonplace third-party advertising tools.  While "Privacy Zuckering" dark patterns are deceptive, it is not clear that companies in our study intended to deceive their users.  Nor is it clear the extent to which these companies were aware how tools are feeding data about users' health information to Facebook as they engage with ads (26).

While tools we identified are not inherently good or bad, applying commonplace advertising tools designed for social media marketing can expose sensitive health information in the form of leads.   These marketing tools reveal a "dark pattern" used to track vulnerable patient journeys across platforms as they browse online, in some ways unclear to the companies and patient populations who are engaging through Facebook.

While the digital medicine ecosystem relies on social media to recruit and build their businesses through advertising-related marketing channels, these practices sometimes contradict their own stated privacy policies and promises to users.  As previously stated, the authors have disclosed findings through proper channels prior to the submission of this work to allow each company time to respond and notify users if a breach occurred.  We hope that the detail around these vulnerabilities inspire deeper introspection into the tools and tactics that PHR companies utilize to increase their reach towards the patients they seek to serve and protect.

# Figures

## Figure 1: Summary of Trackers & 3rd Parties

| Company Name | # of Trackers | Specific Trackers Used | Secondary Vendors | Cross-Site Tracking JSON In Patient Data? | Clear Language in Privacy Policy? | Data Types Identified | Off Facebook Tracking Sample Log | Dark Pattern | Data Shared w/ Facebook |
|---|---|---|---|---|---|---|---|---|---|
| Color Genomics | 3 | https://snap.licdn.com/li.lms-analytics/insight.min.js<br>https://lftracker.leadfeeder.com/lftracker_v1_YEgkB8lGpQWaep3Z.js<br>https://js-agent.newrelic.com/nr-1212.min.js | Google<br>Leadfeeder<br>Nanigans<br>Sprinklr | Yes | No | Content Views | name "color.com"<br>events<br>0<br>id [redacted]<br>type "VIEW_CONTENT"<br>timestamp [redacted]<br>1<br>id [redacted]<br>type "VIEW_CONTENT"<br>timestamp 1577 | Yes | FBCLID |
| Myriad Genetics | 10 | https://www.googleoptimize.com/optimize.js?id=OPT-MD9HP9F<br>https://www.googletagmanager.com/gtag/js?id=UA-[redacted]<br>https://www.googletagmanager.com/gtag/js?id=AW-[redacted]<br>https://script.crazyegg.com/pages/scripts/0075/[redacted].js<br>https://www.googleadservices.com/pagead/conversion.js<br>https://www.googletagmanager.com/gtm.js?id=GTM-[redacted]<br>https://www.googletagmanager.com/gtm.js?id=GTM-[redacted]&l=newDataLayer<br>https://connect.facebook.net/en_US/fbevents.js<br>https://tracker.marinsm.com/tracker/async/[redacted].js<br>https://s.adroll.com/j/roundtrip.js | Google<br>Facebook | Yes | No | Custom Fields<br>Content Views | hereditarycancerquiz.com"<br>events<br>0<br>id [redacted]<br>type "CUSTOM"<br>timestamp 16324507201<br>id [redacted]<br>type "PAGE_VIEW" | Yes | Custom fields<br>FBCLID |
| Invitae | 5 | www.google-analytics.com/analytics.js<br>cdn.evgnet.com/beacon/invitae/engage/scripts/evergage.min.js<br>cdn.pendo.io/agent/static/d96c6792-4d12-453b-64dd-dd74cd6b8f87/pendo.js<br>js-agent.newrelic.com/nr-spa-1177.min.js<br>events.launchdarkly.com/events/bulk/5d4221a4fb53bb072cbd3f12 | Google<br>Launch Darkly<br>Pendo | Yes | Yes | Content Views | ID [redacted]<br>Event VIEW_CONTENT<br>Received on November 17, 2019 at 3:11 PM<br>ID [redacted]<br>Event VIEW_CONTENT<br>Received on November 17, 2019 at 2:48 PM | No | FBCLID |
| Health Union | 8 | https://securepubads.g.doubleclick.net/tag/js/gpt.js<br>https://www.googletagmanager.com/gtm.js?id=GTM-TNDPLXL<br>https://webhooks.fivetran.com/snowplow/da6b9e44-72ae-4a29-9813-2c0488beadc0/com.snowplowanalytics.snowplow/tp2<br>connect.facebook.net<br>geolocation.onetrust.com<br>siteintercept.qualtrics.com<br>sp.analytics.yahoo.com<br>s.yimg.com | Google<br>Fivetran | Yes | No | Content Views | None | Yes | FBCLID |
| Ciitizen | 5 | https://www.googletagmanager.com/gtm.js?id=GTM-PM5678V<br>https://static.hotjar.com/c/hotjar-1853772.js?sv=5<br>https://js.hs-analytics.net/analytics/1641589800000/8259670.js<br>https://js.hs-banner.com/8259670.js<br>https://forms.hsforms.com/embed/v3/counters.gif?key=collected-forms-embed-js-form-bind&count=4 | Google<br>Hotjar | Yes | Yes | Content & Page Views | id [redacted]<br>type "PAGE_VIEW"<br>timestamp 16325967001<br>id [redacted]<br>type "PAGE_VIEW"<br>timestamp 16321686002<br>id [readacted]<br>type "PAGE_VIEW"<br>timestamp 16258494003<br>id [redacted]<br>type "PAGE_VIEW"<br>timestamp 16209727204<br>id [redacted]<br>type "VIEW_CON | | FBCLID |

| | | | | | | TENT"<br>timestamp  157 | | |

Figure 2: Reproduction Steps: Enabling Data to pass Between Digital Medicine Companies and Facebook

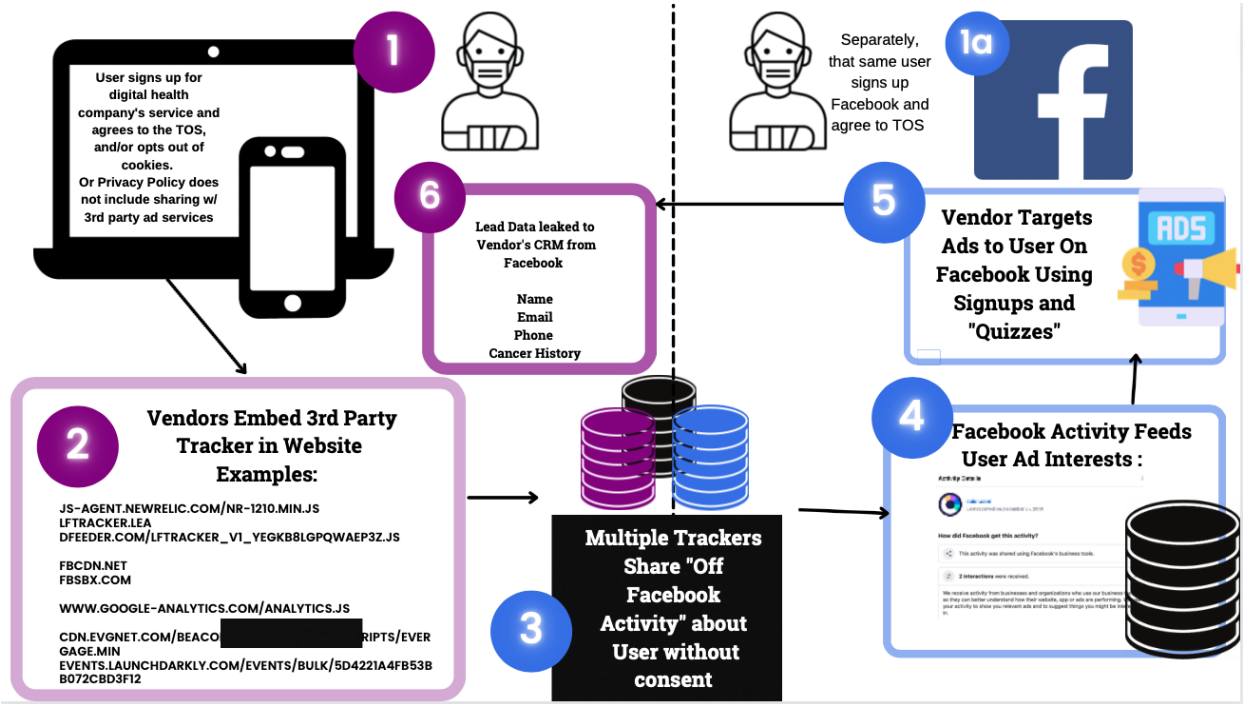

Figure 3: Cross-Site Trackers for Color Genomics

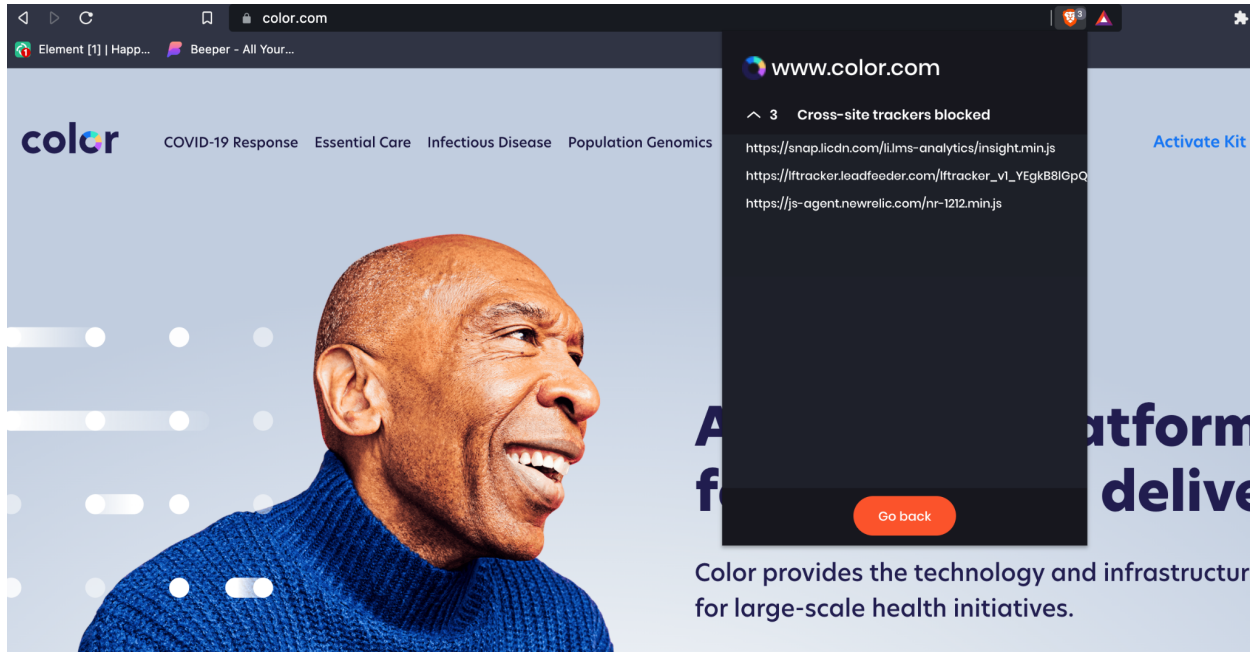

Figure 4:  How Leadfeeder's Service Re-Identifies Patients

# More leads, no extra effort

Sounds too good to be true right? We get it, you've been burned before. All those "One tip to increase leads 5000%" blogs that led you nowhere. Generating leads is difficult.

So we'll tell you real quick how Leadfeeder works.

- Install the Leadfeeder Tracker script on your site
- We identify companies that have visited your website
- We enrich this with an employee contact database
- You send qualified leads directly to your CRM and email

How does this help you? Well, you get to identify companies and decision-makers that are *already* engaging with your content and campaigns.

Figure 5: Sprinkler & Nanigans Unified CXM, Used by Color Genomics

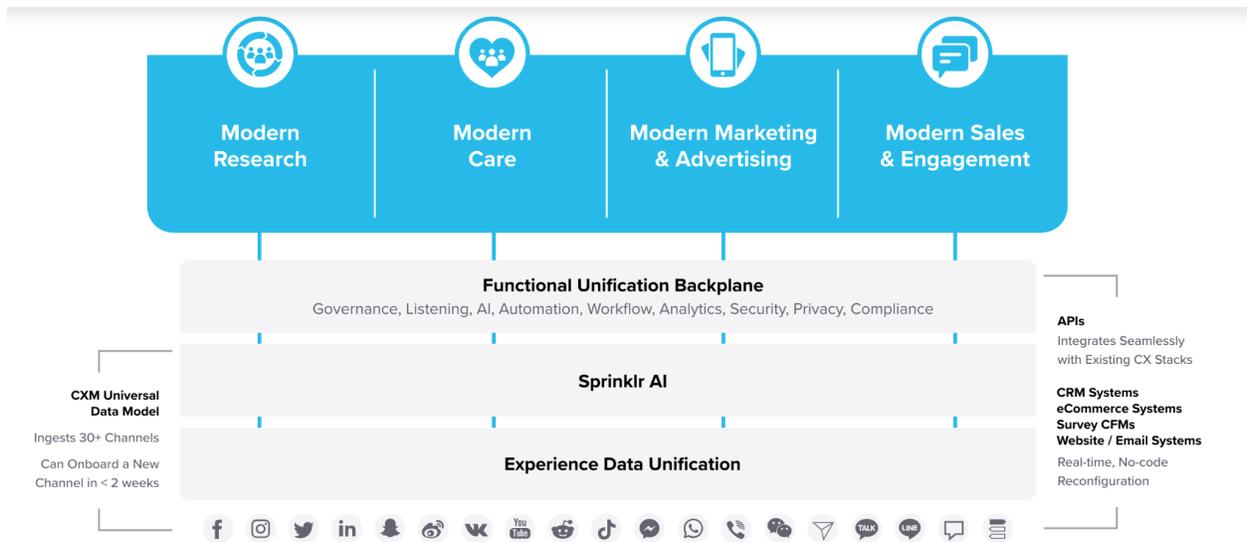

Figure 6: Example messaging to patients from Health Union

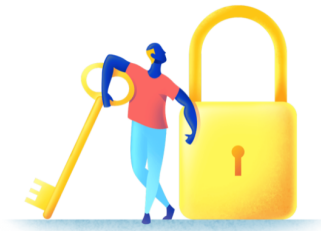

## We take privacy seriously.

We believe people deserve honesty, privacy, and transparency regarding their health and healthcare. We strive to use patient-friendly language and terminology so that the content on our sites is clear and easy to understand for the general population. We always clearly disclose our partnerships and sponsor relationships, and NEVER share or sell any personal identity and contact information (including email addresses) of any community member or individual participant to a sponsor or partner without specific permission to do so.

Figure 7:  Health Union: Contact Walt Disney Company opt-out of the sale of your Personal Information

> **Sale Opt-out.** To submit a request to opt-out of the sale of your Personal Information, you may visit our privacy www.thewaltdisneycompany.com/en/dnsmi/ Rights page or send an email to privacy@health-union.com with the subject line "do not sell info." You have the right not to receive discriminatory treatment for exercising your opt-out right. You may also use an authorized agent to submit a request to opt-out on your behalf if you provide the authorized agent signed written permission to do so. Authorized agents may submit requests using the instructions described above.
>
> **Security**
>
> Health Union uses reasonable administrative, physical and electronic security measures to protect against the loss, misuse and alteration of Personal Information. No transmission of data over the internet is guaranteed to be completely secure. It may be possible for third parties not under the control of Health Union to intercept or access transmissions or private communications unlawfully. While we strive to protect personal information, neither Health Union nor our service providers can ensure or warrant the security of any information you transmit to us over the internet, in particular on open forums. Any such transmission is at your own risk.

Figure 8:  Color's representation to users on Covid19 Testing

> **We do not rent, sell, monetize, or otherwise use any patient data.**
>
> **Color treats samples, data, and information as Protected Health Information.**
> Given the legacy of misuse of genetic information — particularly within communities of color — Color has always been very clear about data, information, and samples. We treat these as Protected Health Information (PHI), which must be safeguarded under the Health Insurance Portability and Accountability Act (HIPAA). As a HIPAA covered entity, Color follows all HIPAA privacy and security rules for safeguarding PHI.
>
> Health data (including your personal information, the fact that you took a test, the test results, and the samples themselves) is all protected. Protected means: secure, private, never sold. Health data is only shared with the explicit permission of the patient.

Figure 9: Example of Cross Site Tracking From Invitae